\documentclass[%
 reprint,
 amsmath,amssymb,
 aps,
 pra,
]{revtex4-1}
\usepackage{color,soul}
\usepackage{subfig}
\usepackage{float}
\usepackage{graphicx}
\usepackage[dvipsnames]{xcolor}
\usepackage{dcolumn}
\usepackage{bm}

\begin{document}

\preprint{APS/123-QED}

\title{Robust Quantum Entanglement Generation and Generation-plus-Storage Protocols with Spin Chains}

\author{Marta P. Estarellas}
\author{Irene D'Amico}
\author{Timothy P. Spiller}
\affiliation{Department of Physics, University of York, York, YO10 5DD, United Kingdom}


\begin{abstract}
Reliable quantum communication/processing links between modules are a necessary building block for various quantum processing architectures. Here we consider a spin chain system with alternating strength couplings and containing three defects, that impose three domain walls between topologically distinct regions of the chain. We show that -- in addition to its useful, high fidelity, quantum state transfer properties -- an entangling protocol can be implemented in this system,  with optional localisation and storage of the entangled states. We demonstrate both numerically and analytically that, given a suitable initial product-state injection, the natural dynamics of the system produces a maximally entangled state at a given time. We present detailed investigations of the effects of fabrication errors, analyzing random static disorder both in the diagonal and off-diagonal terms of the system Hamiltonian. Our results show that the entangled state formation is very robust against perturbations of up to $\sim10\%$ the weaker chain coupling, and also robust against timing injection errors. We propose a further protocol which manipulates the chain in order to localise and store each of the entangled qubits. The engineering of a system with such characteristics would thus provide a useful device for quantum information processing tasks involving the creation and storage of entangled resources.

\end{abstract}

\pacs{Valid PACS appear here}
\maketitle
%

\section{\label{sec:intr}Introduction}

Entanglement is one of the most important quantum phenomenon for quantum technologies and constitutes an essential resource for many applications in quantum information processing, e.g. teleportation protocols \cite{bouwmeester1997,braunstein1998} or one-way quantum computation architectures \cite{raussendorf2001,walther2005}. Moreover, the idea of generating entanglement between two separated qubits is a strategic ingredient for a modular approach to quantum processing architectures \cite{Gualdi2011}. Ion traps \cite{kielpinski2002,monroe2014}, NV-centers \cite{Nemoto2014}, and superconducting qubits \cite{Geller2015} processing architectures require a `quantum communication link' between the modules. Several proposals have been made to design such links \cite{Simon2016,hucul2015,Pfaff2014,Narla2016}. However, all these approaches exploit photons to create and communicate entanglement and, when short distances are involved, the decoherence and losses introduced by having to inter-convert the quantum information between the different physical realisations and flying qubits can lead to inefficiencies and thus low data transfer rates. 
While optical devices and systems are widely regarded as the most applicable candidates for long range quantum communication, spin chains have acquired significant interest within the field of quantum information processing as a means of efficiently transferring information  over short distances \cite{bose2007,kay2010,nikoBook}, and for creating and/or distributing entanglement \cite{spiller2007,sahling2015,Banchi2011,Apollaro2015}. Such model devices can be experimentally implemented for any ensemble of two-level systems where it is possible to engineer the couplings between the systems (sites). Examples include electrons and excitons trapped in nanostructures \cite{damico2007,damico2006,niko2004}, nanometer scale magnetic particles \cite{tejada2001}, or a string of fullerenes \cite{twamley2003}. The use of spins (where 'spin' is used in this wider sense) for both processing and communications avoids some of the pitfalls of using photonics, such as the already mentioned encoding and decoding of information into states of light and the corresponding decoherence and losses associated with these processes.

Spin chains are in general very versatile and can be engineered for different purposes, e.g. to allow for ``perfect state transfer'' (PST, with exactly unit fidelity of transfer) \cite{christandl2004,chris2005,plenio2004,niko2004} or to be tuned to present localised protected states  \cite{estarellas2016,saket2010,saket2013,srivinasa2007}.  Hardware in which the chain defects can be engineered to generate localised protected states are the edges of graphene ribbons \cite{delplace2011}, the edges of honeycomb arrays of microcavity pillars \cite{jacqmin2014}, or Bose-Einstein condensates of $Rb^{87}$ atoms in suitable optical lattice potentials \cite{meier2016,bloch2013}, as well as optical silicon waveguides \cite{Andrea2016}. Previous works have shown spin chains engineered for `PST' to be good candidates for entanglement formation \cite{kay2010} and the knitting of distributed cluster states \cite{ronke2011_2,clark2005,clark2007,yung2005}. Nonetheless, in these proposals to achieve such behaviour requires  all the couplings of the chain to be individually tuned, potentially introducing difficulties in experimentally engineering such systems, and high sensitivity to fabrication errors. 

The introduction of errors, or decoherence, generally has the effect of damaging PST and reducing it to (provided that the damage is not too extensive) high fidelity quantum state transfer (QST). Given that defects and/or decoherence are always present in physical realisations, and given also that protocols exist for error-correction and purification, what is really of interest from a practical and useful device perspective is delivery of high fidelity QST, or generation of high entanglement (close to unity for a pair of qubits). It is this practical and useful approach that forms the focus of our work presented here.

In the first part of the paper, we present a state entangling protocol utilizing a spin chain engineered to have three defects embedded in a dimerized chain with two different coupling strengths. Such defects can be looked at as domain walls between topologically distinct regions of the chain and exhibit topologically localised states. From now on we will refer to these as `topological defects'. 

We first explain the physical model and present both numerical and analytic results for the dynamics and entanglement formation of the system. Clearly, any physical implementation of this protocol will always be subject to errors and imperfections, due to the presence of field fluctuations and fabrication errors, which can be modeled by random static disorder.  We therefore investigate in detail the effects of introducing such fabrication errors. In order to consider different types of disorder, so that our work can be applied to the wide range of physical systems described by the spin chain model, we simulate disorder both on the chain sites and on the couplings, by tuning the diagonal and off-diagonal terms of the Hamiltonian. We show that the entanglement of formation remains very robust against significant levels of off-diagonal disorder and moderate levels of diagonal noise. Robustness of the entanglement against injection timing errors is also demonstrated.

In the second part of the paper we show the possibility of localizing and storing the two entangled qubits in two topologically protected modes, by using a modified version of this chain and performing further transformations to it in an extended version of our protocol. 

Our work represents a new protocol to create and extract and/or store entanglement, that can be potentially implemented in a large range of physical devices, and could be then used towards building efficient quantum communication/processing links.

\section{\label{sec:mod}The Model}

The system considered here is a spin chain of $N=7$ sites with staggered weak ($\delta$) and strong ($\Delta$) couplings, in a distribution such that there are three sites (A, B and C) weakly coupled to the rest of the chain as shown in Fig.\ref{fig1}. We refer to a pair of strongly coupled sites as a ``dimer'' and we term the A, B and C sites as the relevant sites because (i) these present localised eigenstates, (ii) the A and C sites hold the initial state injection, and (iii) the three localised states can be reduced to an effective three-state system which presents a set of particular characteristics, as we further explain later. The coupling energies have been set to be in a $\Delta/\delta=10$ ratio in order to provide both localised states at A, B and C, along with useful overlap between these localised states. This ensures that the system dynamics is propitious for our protocols.  

The spin chain can be described by the following time independent Hamiltonian,

\begin{eqnarray}
\label{hami}
\nonumber{\cal{H}} = \sum_{i=1}^{N}\epsilon_{i}|1\rangle \langle 1|_{i} + \sum_{i=1}^{N-1} J_{i,i+1}[ |1\rangle \langle 0|_{i} \otimes |0\rangle \langle 1|_{i+1} + h.c.],
\end{eqnarray}

with $J_{i,i+1}$ equal to either $\Delta$ or $\delta$ depending on the site (see Fig.\ref{fig1}). The on-site energies, $\epsilon_{i}$, are considered to be site independent (set to zero for convenience) until we add diagonal disorder later on. At any site, in our encoding, a single excitation $|1\rangle$ indicates an ``up" spin in a system initially prepared to have all the spins ``down", $|0\rangle$. In previous literature \cite{estarellas2016,huo2008,Almeida2016} it has been demonstrated that related dimerized chains have high fidelity QST properties, which we will exploit in our protocol.

\subsection{Entanglement generation protocol}
\begin{figure}[h]
\resizebox{0.35\textwidth}{!}{
  \includegraphics{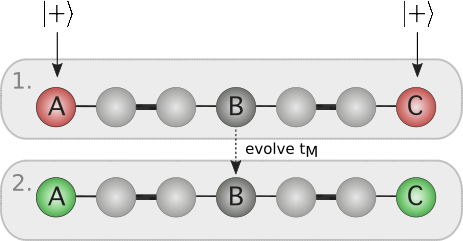}
}
\caption{Entangling protocol: (1) Initial injection of two $|+\rangle=\frac{1}{\sqrt{2}}(|0\rangle+|1\rangle)$ qubit states at site A and C (red) of an ABC type spin chain with $N$=7 sites and alternate couplings, being $\Delta$ the strong coupling (thick line) and $\delta$ the weak coupling (thin line). (2) Time evolution of the system under natural dynamics, until the mirroring time ($t_M$) is reached. At this time qubits A and C are maximally entangled (green).}
\label{fig1}       
\end{figure}

The entanglement generation  protocol presented here is sketched in Fig.\ref{fig1}, and starts with the initial injection at sites A and C of two $|+\rangle=\frac{1}{\sqrt{2}}(|0\rangle+|1\rangle)$ initial states. We then let the system naturally evolve to the mirroring time ($t_{M}$), that is the time needed for an arbitrary initial single-excitation state to propagate to its mirror position in the system. At this time, the injected product state becomes maximally entangled, and the two entangled qubits can be extracted -- if desired -- from sites A and C. 

\section{\label{sec:res}Results}

\subsection{\label{sec:ent}Entanglement creation}

The protocol is initiated at $t=0$, with the injection of two initial $|+\rangle$ states at the chain ends (sites A and C). We can write the initial state in the standard basis as

\begin{eqnarray}	 
|\Psi(0)\rangle &= &\frac{1}{2}\Big(\alpha|0\rangle_A|0\rangle_C+\beta|1\rangle_A|0\rangle_C+\gamma|0\rangle_A|1\rangle_C+ \nonumber\\
& & \kappa|1\rangle_A|1\rangle_C\Big)\otimes|0\rangle_{rest-of-chain}\label{initial}
\end{eqnarray}

In order to quantify entanglement we use the entanglement of formation, $EoF$, a bipartite measure of entanglement for mixed states \cite{wootters1998}. We need to consider the general case of mixed states because although the initial state is pure, due to entanglement with the rest of the chain at later times, the state of just sites A and C is in general mixed (and is calculated from the full chain state by tracing out all the other sites). For a pair of qubits A and C, the $EoF$ is defined by,

\begin{equation}
EoF_{AC}=\xi(C_{AC}),
\end{equation}

being $\xi(C_{AC})=h(\frac{1+\sqrt{1-\tau}}{2})$ and $h=-x\log_2x-(1-x)\log_2(1-x)$ \cite{wootters2001}.
This can be computed by obtaining the square roots of the four eigenvalues, $\lambda_i=\sqrt{\varepsilon_i}$, of the matrix $\rho_{A,C}\widetilde{\rho_{AC}}$ (with $\rho_{AC}$ being the reduced density matrix of sites A and C, and $\widetilde{\rho_{AC}}=(\sigma^A_y\otimes\sigma^C_y)\rho^*_{AC}(\sigma^A_y\otimes\sigma^C_y)$), and arranging these $\lambda_{i}$ in decreasing order. Then $\tau$ is obtained as

\begin{equation}
\tau=(max(\lambda_1-\lambda_2-\lambda_3-\lambda_4,0))^2.
\end{equation}

\begin{figure}[ht!]
\resizebox{0.52\textwidth}{!}{
  \includegraphics{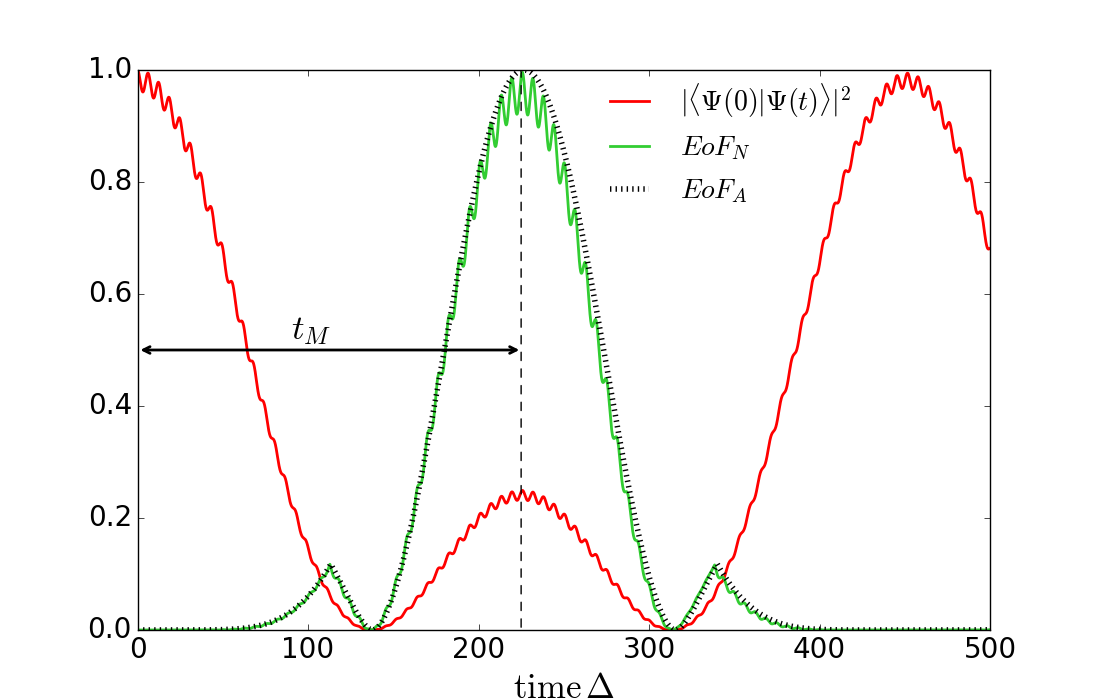}
}
\caption{Fidelity of the initial state $|\Psi(0)\rangle$ (red -dark gray- profile) and numerically calculated $EoF_N$ (green -ligth gray- profile) for a $N$=7 ABC spin chain and $\Delta/\delta=10$. Black dashed profile is the analytically obtained $EoF_A$ of the system.}
\label{fig2}       
\end{figure}

As shown in Fig.\ref{fig2}, the fidelity of the initial state (red profile), which is the probability of recovering the initial overall state as a function of time ($F=|\langle\Psi(0)|\Psi(t)\rangle|^2$), peaks up to unity at twice (and even multiples of) the mirroring time, so at $2t_M$ etc.. In between, at the mirroring time, the two qubits A and C become maximally entangled ($EoF\sim1$) with the following approximated state:

\begin{eqnarray}	|\Psi(t_M)\rangle &\approx &\frac{1}{2}\Big(\alpha|0\rangle_A|0\rangle_C-\beta|1\rangle_A|0\rangle_C-\gamma|0\rangle_A|1\rangle_C- \nonumber\\
& &  \kappa|1\rangle_A|1\rangle_C\Big)\otimes|0\rangle_{rest-of-chain}\label{entangled},
\end{eqnarray}

This behavior can be understood analytically if we consider the Hamiltonian of an effective, reduced `toy model' with just three sites $ABC$ equally coupled ($J_{A,B}=J_{B,C}=\eta$), 

\begin{equation}
H=
\begin{pmatrix}
0 & \eta & 0 \\
\eta & 0 & \eta \\
0 & \eta & 0
\end{pmatrix} \;.
\label{hamiT}
\end{equation}

The trimer eigenstates resulting from the Eq.\ref{hamiT} diagonalization are given by 

	\begin{equation}
	|\phi_{-}\rangle=\frac{1}{2}
	\left(
	\begin{matrix}
	-1\\\sqrt{2}\\-1
	\end{matrix}
	\right)
	|\phi_{0}\rangle=\frac{1}{\sqrt{2}}
	\left(
	\begin{matrix}
	1\\
	0\\
	-1\\
	\end{matrix}
	\right)
	|\phi_{+}\rangle=\frac{1}{2}
	\left(
	\begin{matrix}
	1\\
	\sqrt{2}\\
	1\\
	\end{matrix}
	\right),
	\label{eigenstates}
	\end{equation}
    with $|\phi_{-}\rangle$ having energy $E_{-}=-\sqrt{2} \eta$, $|\phi_{0}\rangle$ having energy, $E_{0}=0$,  and $|\phi_{+}\rangle$ having energy $E_{+}=\sqrt{2} \eta$.

We can write the initial state (Eq.\ref{initial}) in terms of these eigenstates and time evolve each of them through its propagator ($e^{-iEt}$). It is then easy to show that all the terms of our initial state (except the inert $|0\rangle_{A}|0\rangle_{C}$ state) will acquire a $-1$ phase factor, giving the overall state of Eq.\ref{entangled}. 

We wish to analytically characterize the $EoF$ profile. Our overall state for this trimer system at any time can be written as

\begin{widetext}

\begin{equation}	
|\Psi(t)\rangle=\frac{1}{2}\Big(|0\rangle_A|0\rangle_C + \cos(\sqrt{2}\eta t)\big(|1\rangle_A|0\rangle_C+|0\rangle_A|1\rangle_C+|1\rangle_A|1\rangle_C\big)\Big)|0\rangle_B - \frac{i}{\sqrt{2}}\sin(\sqrt{2}\eta t)\Big(|0\rangle_A|0\rangle_C+\frac{1}{2}|1\rangle_A|0\rangle_C+\frac{1}{2}|0\rangle_A|1\rangle_C\Big)|1\rangle_B \; .
\end{equation}

\end{widetext}

We can now find the reduced density matrix for sites A and C by tracing out site B, giving a form

\begin{equation}	
\rho_{AC}=|\alpha_{AC}\rangle \langle\alpha_{AC}| + |\beta_{AC}\rangle \langle\beta_{AC}|
\label{rho}
\end{equation}

where the unnormalised components are given by

\begin{equation}	
\begin{split}
|\alpha_{AC}\rangle = & \frac{1}{2}\Big(|0\rangle_A|0\rangle_C + \cos(\sqrt{2}\eta t)\big(|1\rangle_A|0\rangle_C+|0\rangle_A|1\rangle_C \\
& +|1\rangle_A|1\rangle_C\big)\Big)
\end{split}
\label{comp1}
\end{equation}

and

\begin{equation}
\begin{split}
|\beta_{AC}\rangle = & -\frac{i}{\sqrt{2}}\sin(\sqrt{2}\eta t)\Big(|0\rangle_A|0\rangle_C \\ 
&+\frac{1}{2}|1\rangle_A|0\rangle_C+\frac{1}{2}|0\rangle_A|1\rangle_C\Big) \; .
\end{split}
\end{equation}

At $t_M$, $\sin(\sqrt{2}\eta t)=0$ and $\cos(\sqrt{2}\eta t)=-1$ and Eq.\ref{rho} reduces to the pure state given by Eq.\ref{comp1}, which at this point is maximally entangled due to the additional $-1$ phase factors. We compute the $EoF$ for such an approximated state at all times, by considering $\eta$ as the effective coupling between A-B and B-C of the original $N=7$ site chain. Its value can then be obtained by the eigenvalues immediately above (or below) zero of the overall spectrum and the equation $E_{+}=\sqrt{2}\eta$ ($-E_{-}=\sqrt{2}\eta$).
In terms of the chain parameters, this gives $\eta=\frac{\sqrt{\Delta^2+3\delta^2-\sqrt{\Delta^4+6\Delta^2\delta^2+\delta^4}}}{2}$. After scaling the state dynamics against $\Delta$, we obtain an approximate profile to the full numerical result, as shown by the black dashed line of Fig.\ref{fig2}. Clearly this approximation accurately reproduces the overall entanglement evolution, without though the fine oscillations that are due to the full chain dynamics. 

The mirroring time can also be analytically obtained as  $t_M=\pi/\sqrt{2}\eta$, as it only depends on the effective coupling $\eta$ between the relevant $A,~B,~C$ sites. We thus provide a valid analytic interpretation of our system behavior that demonstrates the importance of the presence of sites A, B and C for the operation of our protocol. 

We also note that the chain length in this model can be increased by adding sets of four sites (two dimers, one either side of site $B$ to preserve the symmetry) and the system still supports the protocol presented here. However, this chain growth would increase the time taken for entanglement creation, exponentially with chain length, due to the exponential decrease of $\eta$ with length.

\begin{figure*}[ht!]
\centering
\resizebox{1\textwidth}{!}{
  \includegraphics{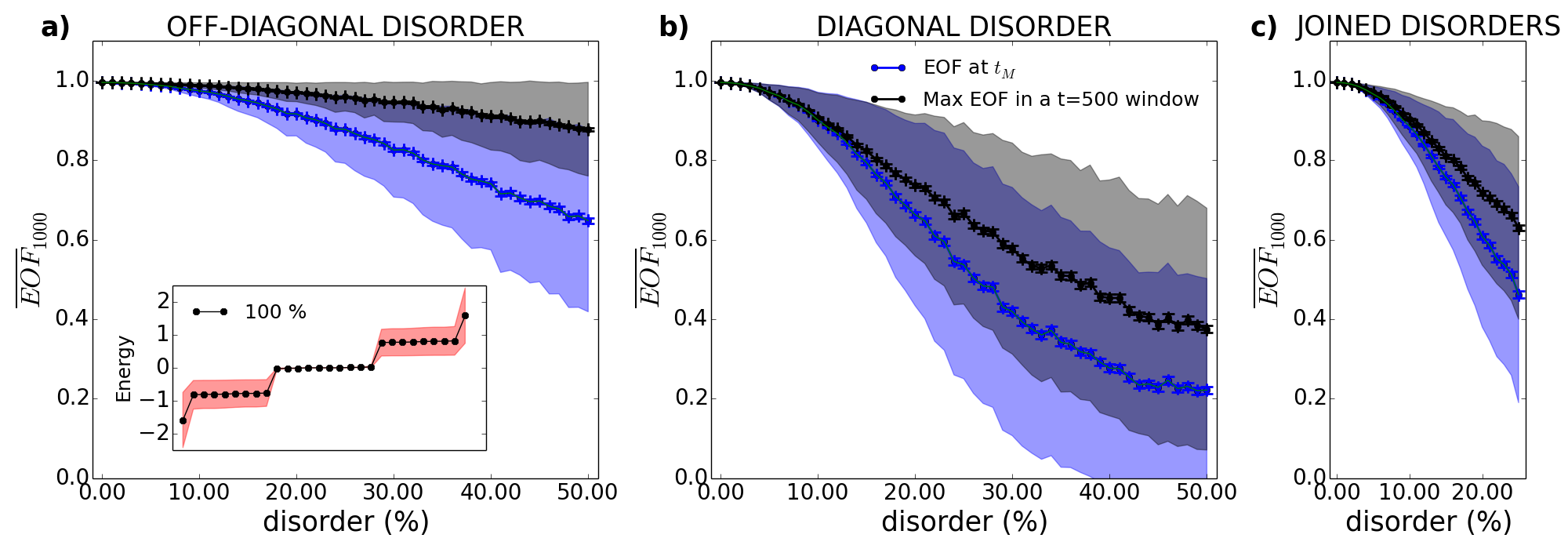}
}
\caption{Averaged EoF at $t_{M}$ (blue -lower- line) and maximum EoF over a 500 units of time window (black -upper- line) for different levels of off-diagonal (a), diagonal (b) and both (c) disorders weighted against the weak coupling ($\delta$). Black and blue shadows represent the standard deviation, black and blue bars represent the standard error of the mean. The inset on the left panel presents the averaged energy spectrum for a chain with a off-diagonal disorder of 100\% the weak coupling (black dots). The standard deviation of each of its eigenvalues corresponds to the red shadow. The dots within the red box correspond to the four exactly zero-energy states.}
\label{fig3}       
\end{figure*}

\subsection{\label{sec:rob}Robustness}

Any practical application of such a protocol will be subject to the presence of fabrication errors in the construction of the device. We have therefore investigated the effects on the entanglement generation of introducing two different types of random static disorder into our system. By considering two different types of such perturbations we aim to simulate a wide range of physical systems subject to different noise. 

The first approach to model local fabrication errors is to consider energy perturbations affecting the sites themselves, and we do this by adding random diagonal disorder to the Hamiltonian. We therefore set $\epsilon_{i}=E·d_{i}·\delta$, where $d_{i}$ is a random number from a uniform distribution between $-1/2$ and $1/2$, and $E$ is a dimensionless parameter that sets the scale of the disorder weighted against the weak coupling $\delta$.

In our second approach to model fabrication errors we consider static coupling errors, by introducing off-diagonal random noise to the Hamiltonian; this is, random disorder in the couplings of the chain. To do so, we set $J_{i,i+1}^{eff}=J_{i,i+1}+E·d_{i}·\delta$, with $J_{i,i+1}\in(\Delta,\delta)$ and $E$ being again a dimensionless parameter weighted against $\delta$ and $d_i$ defined as before.

In order to have an understanding of the practical impact of these two types of fabrication errors on the system, we also consider the case where both disorders are present. We simulate this by adding a randomized perturbation simultaneously to both diagonal and off-diagonal terms of the Hamiltonian.

In the following we compare two scenarios. In the first, given the stochastic nature of these calculations, we present an (ensemble) average over 1000 realisations ($EoF_{1000}$) of the $EoF$ computed at exactly the mirroring time, $t_{M}$, that is expected for the perfect system. Of course in the disordered systems there may well be an error in the actual time at which the $EoF$ peaks; however, in this first scenario we assume that this timing error is unknown and we take the entanglement at the expected time for it to peak, and average this. 

The second scenario corresponds to cases where the timing error could be known in advance, e.g. through calibration of individual devices.  The maximum $EoF$ over a window of 500 units of time is then calculated and, again, the (ensemble) average over 1000 realizations presented. 

To begin we consider results from the first scenario (Fig.\ref{fig3}, blue symbols and shades). These show that the entanglement of the state, at the extraction time $t_{M}$ predicted for the perfect device, and  with disorders smaller than 10\% of the weak coupling $\delta$, is very high ($EoF_{1000}>0.9$) for both types of noise. However, when the disorder levels increase, for the case of diagonal disorder the $EoF$ drops sharply, reaching $EoF_{1000}\sim 0.2$ with a disorder level at 50\% of the weak coupling. Nevertheless, at this same level of off-diagonal disorder the averaged entanglement is still surprisingly high ($EoF_{1000}\sim 0.6$).

The entanglement values generated improve considerably in the second scenario we consider, that is if there is the possibility of making  additional independent measurements on each device (calibration). Given that we are here considering fabrication disorder, this would be equivalent to the  calibrations which are implemented routinely on current electronic components to get their exact specifications. Calibration would enable the state extraction to be performed at the time when the $EoF$ is maximum, for each individual (disordered) device. For this reason there is also significant value in considering the maximum $EoF$ over a time window.
The ensemble average of this entanglement then specifies the average entangling performance on a device selected at random from the ensemble, on the basis that it can then be calibrated for its evolution time scale prior to use. As seen in Fig.\ref{fig3} ((b) panel, black symbols and gray shades), when considering diagonal disorder, the maximum entanglement over a time window of 500 does not go lower than $EoF_{1000}=0.4$ even with noise perturbations at 50\% of the weak coupling. 

For disorder added to the couplings this second scenario is extremely robust, with maximum average entanglement value over the time window always above $EoF_{1000}=0.9$ even for noise perturbations at 50\% of the weak coupling (Fig.\ref{fig3}, (a) panel, black symbols and gray shades).

When both disorders are added (panel (c) of Fig.\ref{fig3}, where we plot up to 25\% disorder), we get a similar trend  to the one obtained with the effect of diagonal disorder only in both $EoF$ measurements. We can conclude from this that diagonal fabrication errors will be dominant for this system, and so reduction of these is the most important practical challenge with regard to fabrication errors. 

The robustness of the system against off-diagonal disorder is remarkably high. This type of noise only affects the upper and lower band of the energy spectrum in a symmetrical way, leaving the genuinely zero-energy states at zero. Once again, this can be understood considering our trimer `toy model' with disorder added to the couplings, such that $\eta+d$ and $\eta+e$ are the off-diagonal terms in the Hamiltonian 
\begin{eqnarray}
\begin{aligned}
H=
\begin{pmatrix}
0 & \eta+d & 0 \\
\eta+d & 0 & \eta+e \\
0 & \eta+e & 0
\end{pmatrix} \; ,
\end{aligned}\nonumber
\end{eqnarray}
which yields the eigenvalues
\begin{equation}
\varepsilon=\pm\sqrt{2\eta^2+2\eta d+2\eta e+d^2+e^2}\;,\;0 \; .
\label{noisetrimer}
\end{equation}

As seen in Eq. (\ref{noisetrimer}), the diagonalization of such perturbed Hamiltonian leaves the zero-energy state undisturbed. 

The same behavior is observed when considering the complete ABC chain system. In the inset of the left panel of Fig.\ref{fig3} we show the energy levels of our $N=7$ site system, up to the two-excitation subspace (and ignoring the inert, zero-excitation, `vacuum' state), and with a level of disorder $E=1$, corresponding to 100\% of the weak coupling $\delta$. The spectrum comprises 7 single-excitation energy states plus $21={N!}/[{2!(N-2)!}]$ two-excitation states. In this case the latter can be expressed approximately as product state combinations of the former. This enables understanding of the 10 states (close to zero energy) sitting in the gap between two `bands'. In the single-excitation subspace the spectrum consists of three states belonging to the relevant $ABC$ sites [see Eq. (\ref{eigenstates})] with energies $\pm\sqrt{2}\eta,0$, sitting in the gap, and two states in the upper `band' and two states in the lower `band'. In the two-excitation subspace the $ABC$ system can be thought of as generating a trimer of three `hole' states (a `hole' being the lack of an excitation in the all excited $ABC$ system and therefore effectively equivalent to the one-excitation states), leading to three states in the gap with energies $\pm\sqrt{2}\eta,0$. Four additional two-excitation states in the gap can be understood analytically as products of a single-excitation state from the upper band with a single-excitation state from the lower band state. The sublattice (or chiral) symmetry of our system imposes mirror symmetry about zero energy on the spectrum; hence, when taking products of single-excitation upper and lower band states we obtain in the gap two exactly zero-energy states and two states with very small energy, equal to the energy difference between the two single-excitation states in a band, which is of the order of $\eta$. We therefore have four exactly zero-energy states (in a rectangular red box in Fig.\ref{fig3}) in the gap, along with three states at very small positive energy and three states at very small negative energy, but still clearly within the gap. As already mentioned, the off-diagonal disorder perturbs the system spectrum symmetrically; consequently the exactly zero-energy states contained in the box in Fig.\ref{fig3} are completely protected. At zero energy, the single-excitation and two-excitation (single `hole') trimer states do not move in energy according to Eq. (\ref{noisetrimer}), while the two single excitation product band states suffer cancelling shifts. 

We can now also understand from this analysis why the $EoF$, even though very robust, does decrease somewhat as the off-diagonal disorder is increased. Despite the four exactly zero-energy states in the middle of the energy gap being protected against this type of noise, the dynamics of the entangling protocol also involves other states in the energy gap (which do suffer small effects due to off-diagonal noise). The states forming the bands also contribute small amplitudes to the protocol dynamics, as the initial states are prepared as site, rather than energy, eigenstates.

\begin{figure}[ht!]
\resizebox{0.45\textwidth}{!}{
  \includegraphics{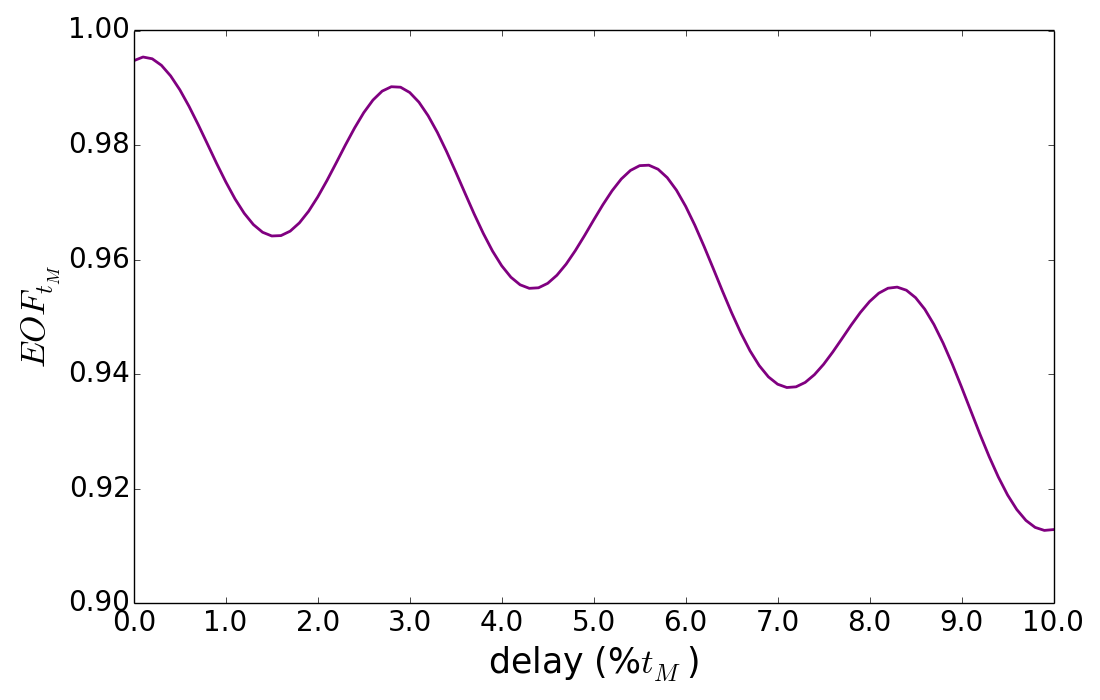}
}
\caption{Entanglement of formation at exactly the mirroring time ($t_M$) against the input delay as a fraction of the mirroring time, for a $N$=7 ABC spin chain with two initial $|+\rangle$ injections at the ends (site A and C).}
\label{fig6}       
\end{figure}

To consider another practical form of error with this model, we also investigate how the asynchronous injection of the two initial $|+\rangle$ states at sites A and C affects the $EoF$ value found at exactly $t_M$. The effect of this error on the protocol is shown in Fig.\ref{fig6}. We observe that even with an injection time delay of 10\% $t_M$, the $EoF$ is still at the high value of 0.91, and for few \% error it is substantially higher. We can conclude that our protocol is therefore also robust against asynchronous state injections.

\begin{figure}[h]
\resizebox{0.45\textwidth}{!}{
  \includegraphics{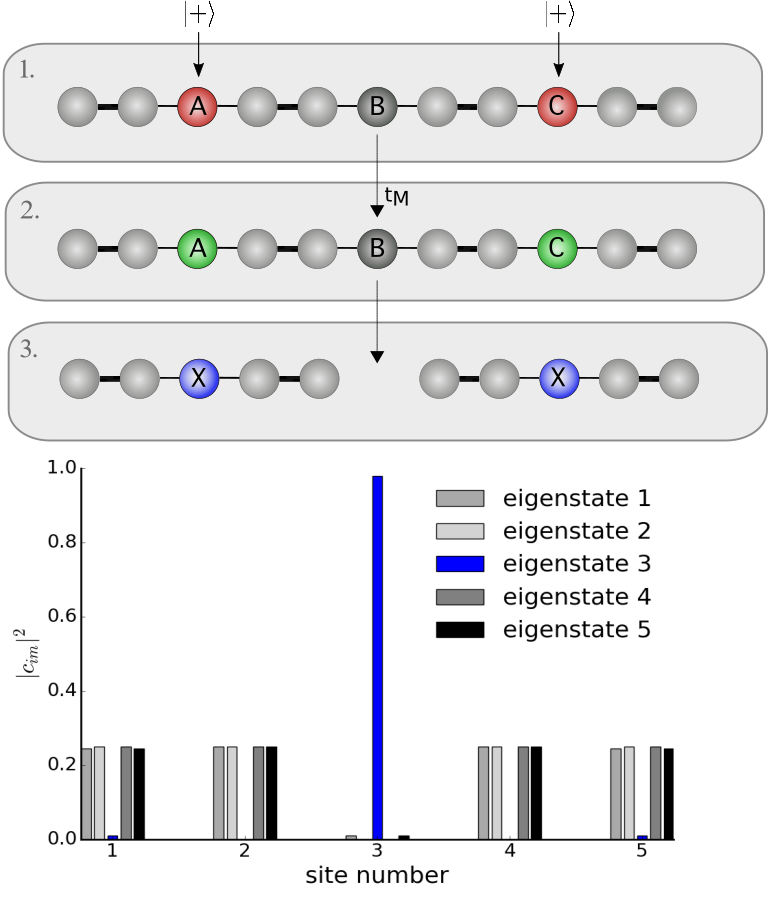}
}
\caption{
Entanglement generation-plus-storage protocol: (1) Initial injection of two $|+\rangle=\frac{1}{\sqrt{2}}(|0\rangle+|1\rangle)$ qubit states at site A and C (red). (2) Time evolution of the system until the mirroring time ($t_M$) is reached: qubit A and C are now entangled (green). (3) At $t_M$, decouple site $B$ completely: the two extangled qubits then localise at the two equivalent sites $X$ in the two now independent chains. The lower panels show the occupation probability distributions for the 5 eigenstates of one of the newly separated chains. The eigenstate peaking at site $X$ is highlighted in blue.}
\label{fig4}       
\end{figure}

\subsection{\label{sec:loc} Entanglement generation-plus-storage protocol}

For quantum processing purposes a very useful facility to control is the production of entanglement and its storage until the rest of the system needs to utilise it.
The $ABC$-type chains indeed offer this flexibility.

To facilitate storage in addition to entanglement generation, we consider a slightly different $ABC$-type chain, with two dimers at the edges of the chain, so that the three $ABC$ defects (solitons) are now completely embedded in the dimerized chain (see Fig.\ref{fig4}).
Following the same entangling protocol as that demonstrated in Fig.\ref{fig2} we obtain essentially identical behavior in both the dynamics and the resilience against disorder. The additional step of the generation-plus-storage protocol incorporates a `switching off' (decoupling) of site $B$ at $t_M$. This separates the system into two independent but equivalent chains, whose single-excitation spectra each contain a topologically protected, strongly localised eigenstate at site $X$ (see the lower panel of Fig.\ref{fig4}). The presence of this state can be explained by considering site $X$ as a defect between two topologically distinct configurations, giving rise to a spatially localised state at zero energy within an energy gap \cite{estarellas2016}. The occupation probability distributions of the five single-excitation eigenstates for the new separated chain are presented in Fig.\ref{fig4}. Note that the middle site contains almost the entire occupation probability of the localised eigenstate (highlighted in blue); this state is basically completely localised at site $X$ with negligible contributions at other sites. 

To model the decoupling we assume that this can be performed on a time scale much shorter than $t_M$ and so employ the sudden approximation to decompose the state of the fully coupled system into the eigenstates of the new decoupled system(s). From our previous asynchronous injection studies we can deduce that the errors caused by time delays on the decoupling of site $B$ will have a minor effect to the overall protocol. We also note that efficient experimental methods to perform similar types of decoupling have been proposed, i.e. applied to molecular spin-chain systems \cite{timco2009}.

\begin{figure}[t]
\resizebox{0.45\textwidth}{!}{
  \includegraphics{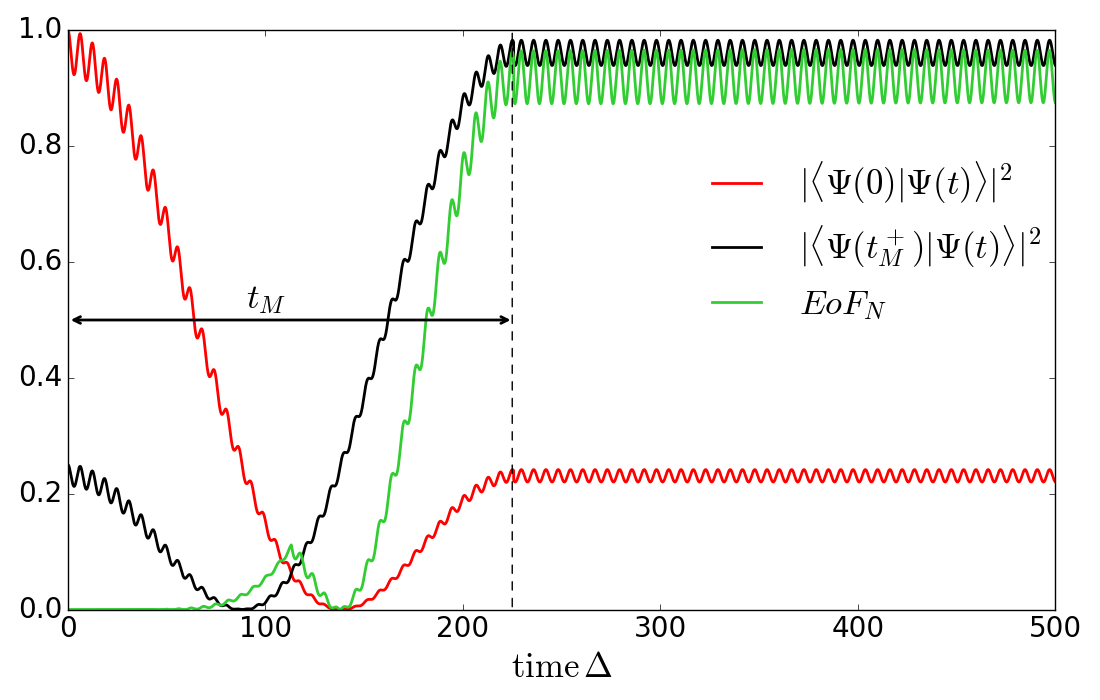}
}
\caption{Overall dynamics of the entanglement generation-plus-storage protocol with the fidelity of the system against the initial state, $|\Psi(0)\rangle$, (red -dark gray- profile) and against the overall entangled state at $t_M$, $|\Psi(t_M)\rangle$,(black profile) along with the numerical calculation of the $EoF_N$ for the whole chain (green -lower light gray- profile).}
\label{fig5}       
\end{figure}

Now at $t=t_M^-$, immediately before decoupling site $B$, the full coupled dynamics generates two entangled qubits that are indeed localised at the sites $A$, $C$, that is the two $X$ sites of the newly decoupled chains at time $t=t_M^+$, immediately after decoupling.  It is therefore expected that the entangled state will inherit the topological localisation of the two shorter and equivalent chains after site $B$ is decoupled.  Indeed this is confirmed by our numerical simulation based on the sudden approximation and shown in Fig.\ref{fig5}. The dynamics of such a protocol show that the fidelity of the overall state at $t_M^+$ ($F=|\langle\Psi(t_M^+)|\Psi(t)\rangle|^2$), once site B has been decoupled, does not reach values lower than 0.9. For extraction purposes, the entanglement will be most useful if it is localised at just the sites $X$ shown in Fig.\ref{fig5}. Therefore we calculate the $EoF$ just for those two sites, by tracing out the rest of the chain. We show that the resultant $EoF$ does not drop below 0.9 either, meaning that the probability of finding the two entangled qubits localised at sites $X$ is basically constant and very high with time after site $B$ is decoupled. This type of localisation is also shown to be extremely protected against disorder, as already shown in previous work \cite{estarellas2016}.

\section{\label{sec:concl}Conclusions}

We have presented a robust entangling gate protocol using spin chains, as well as proposing a protocol to localise and store the two resulting entangled qubits. We have shown numerically and analytically that, after a suitable initial state injection, the natural dynamics of a three-defect, $ABC$-type chain gives rise to the formation of two maximally entangled qubits. These two entangled qubits can be either extracted at a known time $t_M$, or localised and stored, so that the extraction and usage of such a resource can be done at any desired time. The resulting entanglement of formation has been shown to be very robust against two potential fabrication errors of the chain, and also time delay errors on the state injection. We conclude that diagonal errors are more damaging than off-diagonal disorder (against which there is excellent robustness), so in practical implementations diagonal disorder is the fabrication error to focus on reducing. We also conclude that timing injection errors at the few \% (of $t_M$) level also have a very small effect on the performance of the protocol.

The model we have presented is simple, demonstrates high fidelity for quantum state transfer and entanglement generation, and exhibits significant resilience against fabrication errors and timing errors. All this suggests that our proposals provide good candidates for the realisation of reliable quantum communication/processing links between modules in quantum processors and networks. Our approach could be used across the range of physical hardware types that can be mapped onto the spin chain Hamiltonians. It may therefore offer an alternative to the application of optical devices and the inter-conversion of quantum information, for short distance communications.
Our protocols have potential for application in several quantum computer architectures and across of a variety of platforms, particularly where 'off-line', robust, entanglement creation and distribution between two parties is required as a resource. 

\bibliography{papers_bib}

\end{document}